\documentclass[%
 reprint,
 amsmath,amssymb,
 aps,
]{revtex4-2}

\usepackage{float}
\usepackage{graphicx}
\usepackage{dcolumn}
\usepackage{bm}

\usepackage{hyperref}

\begin{document}

\title{The most severe imperfection governs the buckling strength \\ of pressurized multi-defect hemispherical shells}

\author{Fani Derveni}%
\affiliation{%
 Ecole Polytechnique Fédérale de Lausanne (EPFL)\\
    Flexible Structures Laboratory\\
    CH-1015 Lausanne, Switzerland
}%

\author{Florian Choquart}
\affiliation{%
 Ecole Polytechnique Fédérale de Lausanne (EPFL)\\
    Flexible Structures Laboratory\\
    CH-1015 Lausanne, Switzerland
}

\author{Arefeh Abbasi}
\affiliation{%
 Ecole Polytechnique Fédérale de Lausanne (EPFL)\\
    Flexible Structures Laboratory\\
    CH-1015 Lausanne, Switzerland
}

\author{Dong Yan}
\affiliation{%
 Ecole Polytechnique Fédérale de Lausanne (EPFL)\\
    Flexible Structures Laboratory\\
    CH-1015 Lausanne, Switzerland
}

\author{Pedro M. Reis}%
 \email{pedro.reis@epfl.ch}
\affiliation{%
 Ecole Polytechnique Fédérale de Lausanne (EPFL)\\
    Flexible Structures Laboratory\\
    CH-1015 Lausanne, Switzerland
}%


\begin{abstract}
\noindent We perform a probabilistic investigation on the effect of systematically removing imperfections on the buckling behavior of pressurized thin, elastic, hemispherical shells containing a distribution of defects. We employ finite element simulations, which were previously validated against experiments, to assess the maximum buckling pressure, as measured by the knockdown factor, of these multi-defect shells. Specifically, we remove fractions of either the least or the most severe imperfections to quantify their influence on the buckling onset. We consider shells with a random distribution of defects whose mean amplitude and standard deviation are systematically explored while, for simplicity, fixing the width of the defect to a characteristic value. Our primary finding is that the most severe imperfection of a multi-defect shell dictates its buckling onset. Notably, shells containing a single imperfection corresponding to the maximum amplitude (the most severe) defect of shells with a distribution of imperfections exhibit an identical knockdown factor to the latter case. Our results suggest a simplified approach to studying the buckling of more realistic multi-defect shells, once their most severe defect has been identified, using a well-characterized single-defect description, akin to the weakest-link setting in extreme-value probabilistic problems.
\end{abstract}

\maketitle

\section{Introduction}

Shells structures have a wide range of applications in civil and aerospace engineering~\citep{hoff1966thin, godoy_buckling_2016, ferretto2023integrated} and bio-engineering~\citep{lidmar2003virus, yin_stress-driven_2008, katifori2010pollengrains}. Despite their long-recognized benefits for load-bearing capacity and high enclosing volumes, thin shells are prone to catastrophic failure through sub-critical buckling~\citep{pearson2006collapse}. The seminal work by Koiter~\cite{koiter_over_1945} proposed a foundational theory of elastic stability to analyze the post-buckling behavior of these structures. However, classical theoretical predictions~\citep{Zoelly1915} for the buckling onset systematically overestimate actual measurements. This mismatch prompts the introduction of the knockdown factor, $\kappa\leq 1$; the ratio between the actual critical buckling pressure of realistic/imperfect shells and the equivalent theoretical prediction for perfect ones. The imperfection sensitivity of shells~\citep{von1939buckling,von1940influence,hutchinson1970postbuckling}, arising from variations in thickness and loading or the presence of geometric defects, poses a significant challenge in designing and accurately predicting their buckling behavior. 
To reconcile theory with practice in engineering applications requiring lightweight yet strong structures, advanced design and analysis techniques are essential for optimizing knockdown factors while minimizing mass and striving for buckling-induced functionalities~\citep{reis_perspective_2015,shim2012buckling,terwagne2014smart,yang2024complex}.

Past research on shell buckling has historically been driven by theoretical and computational efforts~\citep{bijlaard1960elastic, hutchinson_effect_1971, budiansky1972buckling, hutchinson2016buckling}, with experimental work often taking a second stage. Recent experimental advancements have revitalized the field, thanks to a rapid and precise coating technique~\citep{lee_fabrication_2016} to produce nearly uniform thickness. These shells exhibit lower susceptibility to buckling under pressure than previous fabrication methods, such as metal spinning~\citep{kaplan_nonlinear_1954, homewood1961experimental} or plastic vacuum drawing~\citep{seaman1961nature}. Even more importantly, this experimental technique can be adapted to producing shells with meticulously designed imperfections~\citep{lee2016geometric, abbasi2021probing, yan2021magneto}. Subsequent studies have primarily focused on single imperfections localized at the shell pole via experiments and numerical simulations, including constant thickness geometrical imperfections, such as dimples~\citep{thompson_probing_2017,jimenez_technical_2017,marthelot_buckling_2017} and bumps~\citep{abbasi2023comparing}, or thickness variations~\citep{yan2020buckling}. 

Extending beyond the single-defect case, Derveni et al.~\cite{derveni2023defect} explored the effect of interactions between defects by investigating spherical shells with two imperfections and unveiled a defect-defect interaction regime. Within this interaction regime, the knockdown factor can be either enhanced or reduced, the extent of which is dictated by the critical buckling wavelength of the shell~\citep{Hutchinson1967}. Past the threshold separation for this interaction, the buckling capacity of two-defect shells closely resembled that of equivalent single-defect shells, suggesting the dominance of their strongest defect.
Two-defect cylindrical shells have also been previously investigated~\citep{wullschleger2006numerical, fan2019critical}.

Developing shells that are less susceptible to buckling requires precision manufacturing. Carlson et al. \cite{Carlson1967} advanced the production of metallic spherical shells through electroforming, significantly improving specimen quality via a chemical polishing treatment. Testing these spheres revealed a remarkable increase in the knockdown factor from $\kappa\approx0.05$ to $\kappa\approx0.86$ as severe defects were progressively eliminated. However, this technique also had limitations: it could not systematically vary the distribution of imperfections and was time-consuming. There is a need for novel techniques to investigate more realistic shells with a well-defined distribution of imperfections. 

The buckling of cylindrical shells with multiple imperfections has been tackled with various probabilistic methods, such as modified truncated hierarchy~\citep{amazigo1969buckling} or Monte Carlo~\citep{elishakoff1982reliability,elishakoff1985reliability}. However, considerably less attention has been given to spherical multi-defect shells. Recently, Derveni et al.~\cite{derveni2023probabilistic} examined the case of spherical shells with a random distribution of geometric imperfections on the surface of the shells. That work demonstrated that when the amplitude of the defects is sampled from a lognormal distribution, the resulting knockdown factor can be described using a 3-parameter Weibull distribution. This observation categorized shell buckling as part of a broader group of statistical phenomena known as extreme-value statistics~\citep{jayatilaka_statistical_1977, le2015modeling}. Subsequent stochastic analyses through the weakest-link model~\citep{baizhikova2024uncovering} provided a novel theoretical framework for systematically reevaluating defect severity more efficiently and effectively than the experiments mentioned above by Carlson~\cite{Carlson1967}.
Quantifying how removing specific localized defects influences buckling strength in multi-defect shells is yet to be fully addressed.

Here, we investigate how the buckling strength, as measured by the knockdown factor, of a hemispherical shell containing a distribution of defects is modified by the systematic removal of localized imperfections in order of severity. We seek to establish the relationship between the knockdown factor and the maximum amplitude defect. We will build upon previous numerical studies by Derveni et al.~\cite{derveni2023defect,derveni2023probabilistic} and provide context for the landmark experiments of Carlson~\cite{Carlson1967}. Finite element simulations for multi-defect shells will be compared to equivalent single-defect shells containing the most severe (\textit{i.e.}, the \textit{worst)} defect sampled from the distribution. Our main finding is that the imperfection with the largest amplitude dictates the buckling response, and upon its removal, the knockdown factor increases. This finding provides conclusive quantitative evidence of the significance (observed and studied widely in past literature) of the most severe defect in governing shell buckling and, moreover, in generating stronger shells upon controlled defect removal.

Our paper is organized as follows. First, in Sec.~\ref{sec:probdefinition}, we define the research problem at hand. Next, in Sec.~\ref{sec:FEM}, we describe the methodology of our simulations based on the Finite Element Method (FEM). In Sec.~\ref{sec:Weibull}, we present FEM results 
on the knockdown factor statistics. Results from a systematic exploration of removing increasing fractions of defects from small-to-large amplitudes are presented in Sec.~\ref{sec:smalltolarge} and from large-to-small in Sec.~\ref{sec:largetosmall}. Comparisons of the buckling strength of these multi-defect shells with single-defect shells are provided in Sec.~\ref{sec:FEMsingledefect}. Finally, in Sec.~\ref{sec:conclusions}, we summarize the conclusions of our work and offer suggestions for future research. 

\section{Problem Definition}
\label{sec:probdefinition}

We consider a shell geometry identical to that of Derveni et al.~\cite{derveni2023probabilistic}, which is summarized next for completeness. Specifically, our investigation focuses on thin, elastic, hemispherical shells of radius, $R=25.4~\mathrm{mm}$, and thickness, $t=0.23~\mathrm{mm}$, containing multiple localized geometric imperfections ($N \gg 1$) distributed randomly on their surfaces. The hemispherical shells are clamped at their equator. The radius-to-thickness ratio is $\eta=R/t=110$. These specific values of $R$ and $t$ were chosen to match and build upon a series of past studies~\citep{lee2016geometric,marthelot_buckling_2017,yan2020buckling, abbasi2021probing, derveni2023defect, derveni2023probabilistic}, involving a combination of experiments and FEM simulations, with different $R/t$ values having been explored \cite{jimenez_technical_2017, derveni2023defect}.
\begin{figure}[b!]
    \centering
    \includegraphics[width=0.78\columnwidth]{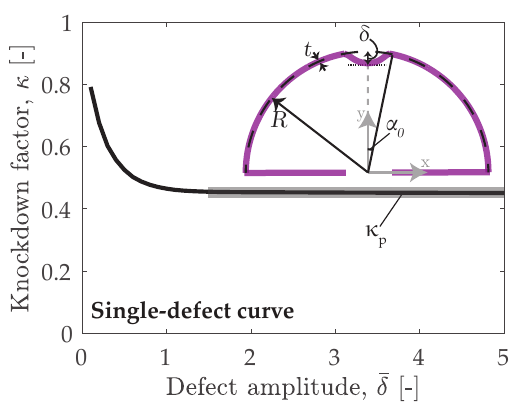}
    \caption{Knockdown factor, $\kappa$, versus normalized defect amplitude, $\overline{\delta}$, for a shell with a \textit{single defect} at its pole. The black line presents FEM data that was validated previously against experiments \citep{lee2016geometric} for $\lambda=1$. The boxed gray region corresponds to the plateau knockdown factor, $\kappa_\mathrm{p}$. Inset: Schematic diagram of the axisymmetric geometry of this single-defect shell, labeled with the relevant geometric quantities.}
    \label{fig:onedefect}
\end{figure}
\begin{figure*}[t!]
    \centering
    \includegraphics[width=0.83\textwidth]{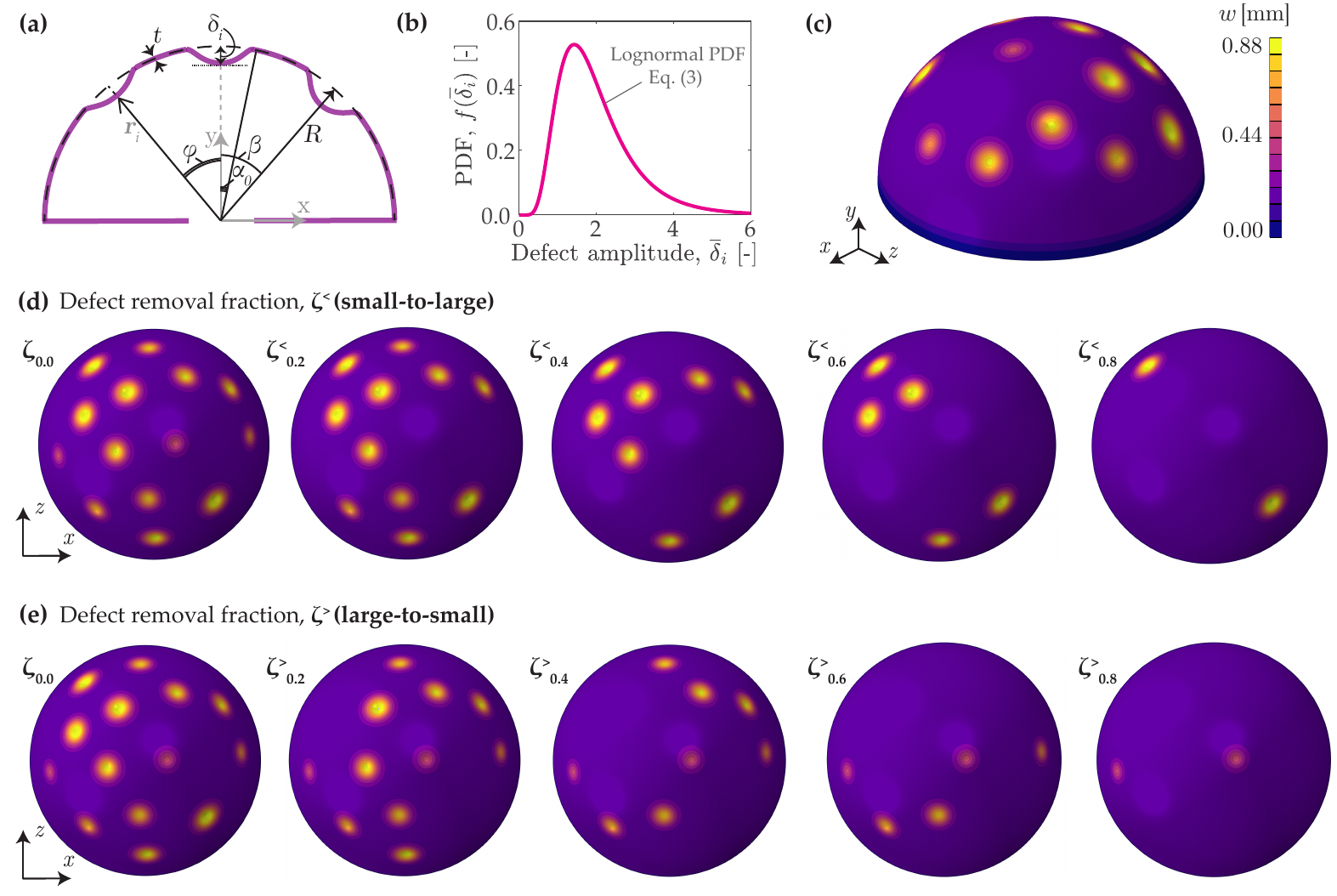}
    \caption{Problem definition and reference geometry of shells containing multiple defects. (a) 2D schematic of a hemispherical shell defining all the relevant geometric variables. (b) Probability density function, $f(\overline{\delta}_i)$, of lognormally distributed defects with the defect amplitude, $\overline{\delta}_i$, as the random variable. (c) 3D representation of a shell with $\varphi_\mathrm{min}=25^{\circ}, ~\overline{\delta}=2.0$ and $\Delta\overline{\delta}=1.0$; the colorbar indicates radial deviation from a perfect shell, $w$. (d, e) Top views in the $x-z$ plane of a series of shells: defects are systematically removed from each shell from (d) small-to-large defect amplitude, $\zeta^<$, or (e) large-to-small defect amplitude, $\zeta^>$. Selected cases of a removal fractions $\zeta^<=\zeta^>=\{0.2, 0.4, 0.6, 0.8\}$ are presented as examples.}
    \label{fig:geometry}
\end{figure*}
Figures~\ref{fig:onedefect} and~\ref{fig:geometry} show representative shells with $N = 1$ and $N \gg 1$ defects, respectively, all shaped as Gaussian dimples with a radial deviation of: 
\begin{align}\label{eqn:geom_gaussiandimple}
w_i(\alpha) = -\delta_i e^{-(\alpha/\alpha_0)^2},
\end{align}
where the subscript $i$ refers to each imperfection of amplitude $\delta_i$ and half-angular width $\alpha_0$, and the variable $\alpha$ is the local angular distance (spherical coordinate) from the center of the imperfection. Similarly to previous studies on shells with a single defect at the pole~\citep{lee2016geometric,jimenez_technical_2017}, throughout the manuscript, we use the following standard dimensionless variables when referring to the amplitude, $\overline{\delta}_i=\delta_i/t$, and width, $\lambda=[12(1-\nu^2)]^{1/4}\,(R/t)^{1/2}\,\alpha_0$~\citep{kaplan_nonlinear_1954} of each defect ($\nu$ is the Poisson's ratio). 

The imperfect shells are pressurized until buckling at the maximum pressure, $p_\mathrm{max}$. As customary in the literature~\citep{seide1960development, peterson1968buckling}, we quantify the buckling strength of the \textit{imperfect} shell by the knockdown factor,
\begin{equation}
    \kappa=\frac{p_\mathrm{max}}{p_\mathrm{c}},
\end{equation}
where $p_\mathrm{c}= 2E\eta^{-2}/\sqrt{3(1-\nu^2)}$ is the classic theoretical prediction for the buckling pressure of the corresponding \textit{perfect} shell geometry~\citep{Zoelly1915}. 

In Fig.~\ref{fig:onedefect}, we present a representative curve obtained from FEM simulations for the knockdown factor, $\kappa$, versus the normalized defect amplitude, $\overline{\delta}$, for a shell containing a \textit{single} defect of dimensionless width $\lambda=1$; see schematic diagram in the figure inset. These data were reproduced from Derveni et al.~\cite{derveni2023probabilistic}, where they were validated against experimental results from Lee et al.~\cite{lee2016geometric}. For large values of $\overline{\delta}$, $\kappa$ reaches a constant plateau, $\kappa_\mathrm{p}=0.45$. The single-defect curve will become important later in our manuscript when interpreting the knockdown factor results for shells with multiple defects. 

Next, we describe the case of imperfect shells containing a \textit{distribution} of localized defects, the central focus of the present study, following a geometry and construction identical to that introduced in Derveni et al.~\cite{derveni2023probabilistic}; see Fig.~\ref{fig:geometry}(a). Each of the $i$-th defect is located at $\beta_i$ and $\theta_i$, the global zenith (polar) and azimuthal spherical coordinates, respectively. The defects are distributed randomly in a spherical cap within $\beta=60^\circ$ so as to 
avoid boundary effects near the clamped equator. Moreover, the center-to-center angular separation between two adjacent defects, $\varphi$, is enforced to have a minimum value of $\varphi_\mathrm{min}$ to prevent overlapping and defect-defect interactions~\citep{derveni2023defect}. 
The amplitude of each $i$-th imperfection is sampled from a lognormal distribution (with a mean amplitude $\langle\overline{\delta}\rangle$ and standard deviation $\Delta\overline{\delta}$) given by the probability density function (PDF):
\begin{align} \label{eqn:lognormal}
    f(\overline{\delta}_i)=\frac{1}{\overline{\delta}_i \sigma \sqrt{2\pi}}
    \exp\left( 
    -\frac{(\ln \overline{\delta}_i -\mu)^2}{2 \sigma^2}
    \right),
\end{align}
where $\mu$ and $\sigma$ define the mean defect amplitude, 
\begin{align} \label{eqn:PDFmean}
    \langle\overline{\delta}\rangle=\exp\left( \mu + \frac{\sigma^2}{2}\right),
\end{align}
and its standard deviation,
\begin{align} \label{eqn:PDFstd}
    \Delta\overline{\delta}=\biggl\{  \left[\exp(\sigma^2)- 1 \right]\, \exp\left( 2\mu + \sigma^2\right)\biggl\}^{1/2}.
\end{align}
In Fig.~\ref{fig:geometry}(b), we provide an example of $f(\overline{\delta}_i)$ for a representative shell design using Eq.~(\ref{eqn:lognormal}) to seed the defects. Throughout, we will enforce that the minimum angular separation between two neighboring defects is $\varphi_\mathrm{min}= 25^\circ$ to avoid defect-defect interactions~\citep{derveni2023defect}. Fig.~\ref{fig:geometry}(c) presents an example for a specific shell design with $(\langle\overline{\delta}\rangle, \Delta\overline{\delta}) = (2.0, 1.0)$.

Our previous studies~\citep{derveni2023defect,derveni2023probabilistic} indicated that, in the absence of defect interactions, the most severe imperfection governs the buckling of spherical shells. However, those studies were not systematic in exploring the geometric parameter space. Furthermore, we did not directly measure the magnitude of the most severe defect in each shell, $\overline{\delta}_\mathrm{max}$ (\textit{i.e.}, the maximum amplitude randomly sampled from $f(\overline{\delta}_i)$), nor did we quantify its effect on the knockdown factor.

In the present study, we will generate multi-defect shell designs with geometric parameters ($\langle\overline{\delta}\rangle, \Delta\overline{\delta}$) and systematically remove fractions of imperfections $\zeta {\in }[0.1, 0.9]$ based on their amplitudes by excluding either (i) the least severe defects (see Fig.~\ref{fig:geometry}d), or (ii) the most severe defects (see Fig.~\ref{fig:geometry}e). We will then conduct FEM simulations on these designs to compute the corresponding knockdown factors. 
Hereon, when we refer to the defect removal fraction, we will use the notation $\zeta^<$ and $\zeta^>$ for \textit{small-to-large} and \textit{large-to-small} defect removal, respectively. 
In Figs.~\ref{fig:geometry}(d,e), for both cases of increasing $\zeta^<$ or $\zeta^>$, we show ($x$-$z$ plane) the initial geometry of shells in a sequence of defect removal fractions of $\{\zeta^<,\zeta^>\}=\{0.2,\,0.4,\,0.6,\,0.8\}$, with the colorbar representing the radial deflection, $w$, away from the perfect spherical geometry. 

\begin{figure}[h!]
    \centering
    \includegraphics[width=0.72\columnwidth]{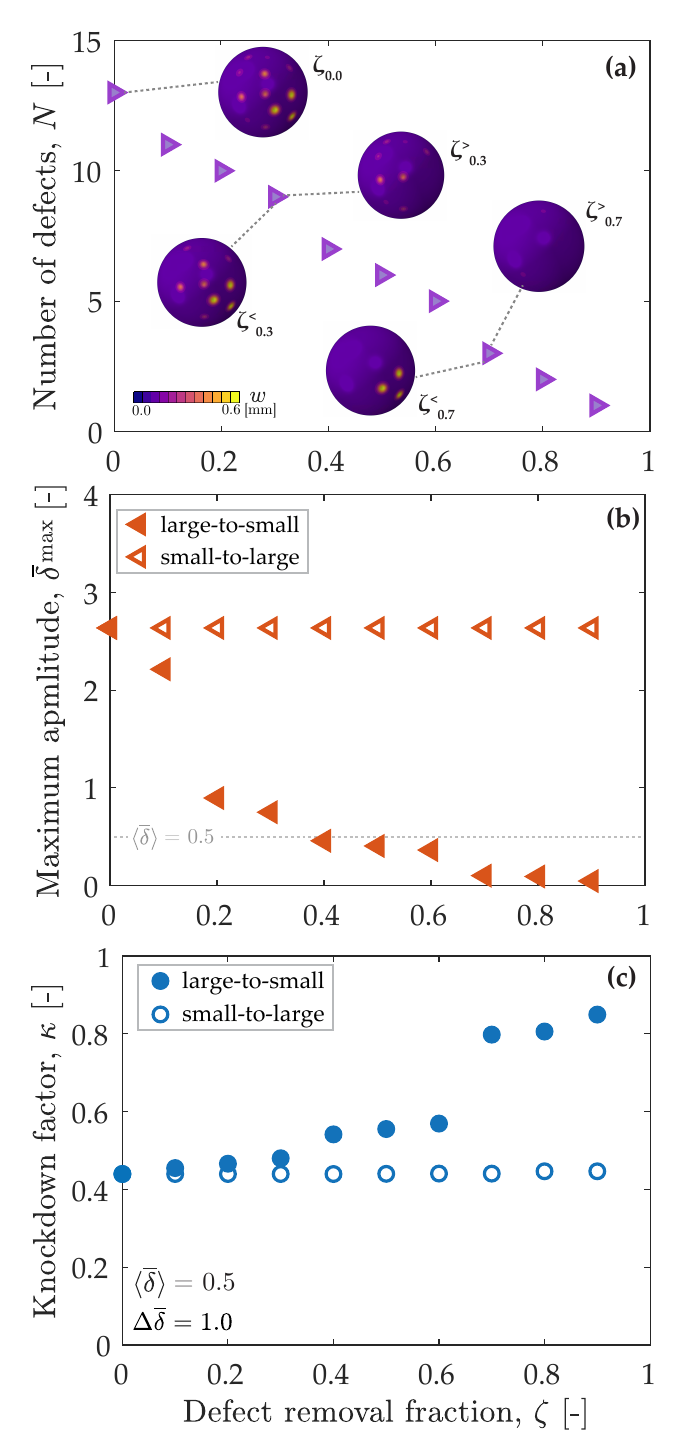}
    \caption{Representative (input) design parameters (a,b) and measured (output) knockdown factor (c) for a single realization ($n=1$) of a multi-defect shell with $(\overline{\delta},\, \Delta\overline{\delta})=0.5,\,1.0)$. (a) Number of defects, $N$, versus defect removal fraction, $\zeta$, for $\varphi_\mathrm{min}=25^\circ$; insets show top views of the shell for selected defect removal fractions; large-to-small, $\zeta^>$, or small-to-large, $\zeta^<$. (b) Maximum amplitude, $\overline{\delta}_\mathrm{max}$, versus $\zeta^>$ or $\zeta^<$. (c) Knockdown factor, $\kappa$, versus $\zeta^>$ or $\zeta^<$.} 
    \label{fig:simpleproblem}
\end{figure}

In Fig.~\ref{fig:simpleproblem}, we present results for a single shell realization ($n=1$) for the two defect-removal scenarios mentioned above: either increasing $\zeta^>$ or $\zeta^<$. This specific example with defects characterized by $(\langle\overline{\delta}\rangle,\Delta\overline{\delta})=(0.5,\,1.0)$, involves a shell containing $N=13$ defects (when $\zeta_{0.0}$). Fig.~\ref{fig:simpleproblem}(a) shows $N$ decreasing linearly as $\zeta^>$ or $\zeta^<$ increase, reaching $N=1$ when $\{\zeta^>,\zeta^<\}=0.9$. Various insets provide top-view visualizations of the corresponding shells.
In Fig.~\ref{fig:simpleproblem}(b), we then plot the maximum defect amplitude, $\overline{\delta}_\mathrm{max}$, for each of the configurations in panel (a). We find that $\overline{\delta}_\mathrm{max}$ decreases as $\zeta^>$ increases (closed symbols) but remains constant with $\zeta^<$ (open symbols). Next, in Fig.~\ref{fig:simpleproblem}(c), we plot the knockdown factor for the shells characterized in panels (a) and (b);
the filled and open circle symbols represent the $\zeta^>$ and $\zeta^<$ data sets, respectively, with qualitatively distinct behavior between the two cases. The knockdown factor of these shells increases from $\kappa=0.44$ at $\zeta^>=0$ to $\kappa=0.93$ at $\zeta^>=0.9$, whereas $\kappa$ remains constant across the full $\zeta^<$ range. Also, in light of Fig.~\ref{fig:simpleproblem}(b), these results suggest that $\kappa$ is related to $\overline{\delta}_\mathrm{max}$, motivating a more comprehensive exploration of the parameter space.

The above results on the dependence of $\kappa$ on $\overline{\delta}_\mathrm{max}$ for a \textit{single realization} of an imperfect shell lead us to establish the two central questions we seek to address: 
\textit{How does the gradual removal of imperfections in a multi-defect hemispherical shell affect the statistics of the knockdown factor? What is the role of the most severe imperfection in setting the buckling strength?}

\section{Methodology: FEM Simulations} 
\label{sec:FEM}

We utilized a computational pipeline based on the Finite Element Method (FEM), implemented in the commercial software ABAQUS/Standard~\citep{Abaqus:2019}. These FEM simulations were 
adapted from our previous work~\citep{derveni2023probabilistic, derveni2023defect}, where we validated them against physical experiments~\citep{derveni2023probabilistic}, and are summarized next for completeness.

We modeled 3D hemispherical shells using S4R shell elements with reduced integration points. First, we generated perfect hemispherical shells and then introduced imperfections according to Eq.~(\ref{eqn:geom_gaussiandimple}) through nodal displacements based on various sets of geometric parameters ($\langle\overline{\delta}\rangle, \Delta\overline{\delta}, \lambda, \varphi_\mathrm{min}$). Each quarter of the shell was discretized with 150 elements in both the meridional and azimuthal directions, resulting in a total of 67\,500 elements in the full hemisphere.

The FEM simulations used a static Riks solver,
selecting an initial arc length increment of $0.1$, a minimum increment of $10^{-5}$, and a maximum increment size of $0.5$. Clamped boundary conditions were imposed at the shell equator, and uniform live pressure was applied to its outer surface. The radius-to-thickness ratio was kept fixed at $\eta = R/t = 110$ (by fixing the radius and thickness of the shell to $R=25.4\,$mm and $t=0.23\,$mm, respectively). The material was assumed to be a neo-Hookean solid with a Young's modulus of $E = 1.26\,$MPa, and a Poisson's ratio of $\nu=0.5$ (assuming incompressibility), so as to represent vinylpolysiloxane (VPS-32, Elite Double 32, Zhermack), consistently with previous experiments~\citep{lee2016geometric,marthelot_buckling_2017, derveni2023defect}. Geometric nonlinearities were taken into account throughout the simulations. 

We explored the geometric parameter space within the following ranges: the lognormal-distributed defects (cf. Eq.~\ref{eqn:lognormal}) had a mean amplitude of $\langle\overline{\delta}\rangle=\{0.5,\,1, \,2\}$, with a standard deviation of  $\Delta\overline{\delta}=\{0.1,\, 0.5,\, 1\}$, and the defect-removal fractions were $\zeta^> {\in }[0.0, 0.9]$ or $\zeta^< {\in }[0.0, 0.9]$, in steps of $0.1$. 
The defects had a fixed normalized width of $\lambda=1$ and were distributed with the minimum angular separation of $\varphi_\textrm{min}=25^\circ$ (non-interacting defects; see Ref.~\cite{derveni2023defect}). For each set of (input) geometric parameters, we generated $n=100$ realizations of statistically equivalent shell designs and analyzed the (output) buckling statistics, as measured by the knockdown factor. 

\section{Knockdown factor probability density functions} 
\label{sec:Weibull}

We start by examining the buckling statistics of shells containing lognormally distributed imperfections, similar to Derveni et al.~\cite{derveni2023probabilistic}, but now focusing on the systematic removal of increasing fractions of imperfections in our multi-defect shells, from the smallest to the largest, and vice versa.
We seek to attest whether the FEM data of the knockdown factor statistics, for a selection of representative cases from the geometric parameter space, can still be described by a 3-parameter Weibull probability density function (PDF)~\citep{weibull1951statistical} when a fraction of the defects are removed. We point the reader to Ref.~\cite{derveni2023probabilistic}, especially its Section 7 and Eq.~(7.1), for details on using this Weibull description for the shell buckling problem at hand, including the functional form of the 3-parameter Weibull PDF, $f(\kappa)$.  
\begin{figure}[h!]
    \centering
    \includegraphics[width=0.68\columnwidth]{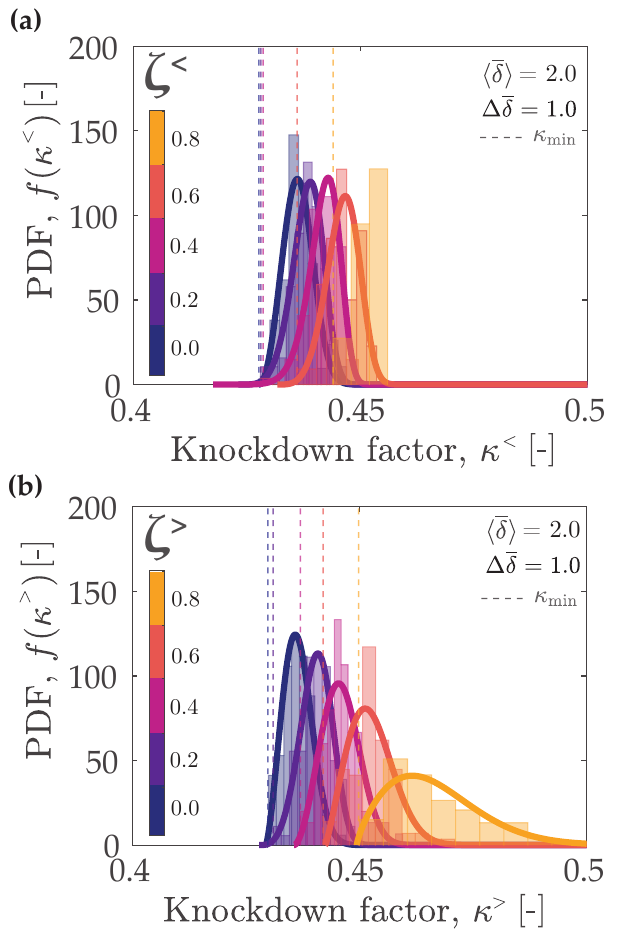}
    \caption{Probability density functions (PDFs), $f(\kappa)$, of the knockdown factor, $\kappa$, for imperfect shell with $(\langle\overline{\delta}\rangle,\,\Delta\overline{\delta})=(2.0,\,1.0)$ for several cases of (a) small-to-large defect removal fractions, $\zeta^<$, and (b) large-to-small defect removal fractions, $\zeta^>$. The colored histograms and the solid lines represent, respectively, the FEM data and the corresponding fits to the 3-parameter Weibull ~\citep{weibull1951statistical} PDF -- cf. Eq.~(7.1) in Ref.~\cite{derveni2023probabilistic} -- for different values of $\zeta^<$ and $\zeta^>$ (see adjacent color bars). Note that the PDF for $\zeta^< = 0.8$ is not shown because $\kappa^<$ was nearly constant across the $n=100$ realizations, and it was difficult to obtain satisfactory fits. The vertical dashed lines indicate the minimum knockdown factor, $\kappa_\mathrm{min}$, of each defect removal scenario.}
    \label{fig:weibullstatistics}
\end{figure}

In Fig.~\ref{fig:weibullstatistics}, we present the FEM-computed statistical results for the knockdown factor of imperfect shells with $(\langle\overline{\delta}\rangle,\,\Delta\overline{\delta})=(2.0,\,1.0)$; panel (a) for the small-to-large ($\zeta^<$) case and panel (b) for the large-to-small ($\zeta^>$) case. In both panels, the histograms correspond to the FEM data, the solid lines represent the Weibull-fitted PDFs, $f(\kappa^<)$ (cf. Eq.~7.1 in Ref.~\cite{derveni2023probabilistic}), and the colorbar represents the removal-fraction values. In Fig.~\ref{fig:weibullstatistics}(a), as increasing fractions of the least severe imperfections are removed, $f(\kappa^<)$ become slightly narrower, but the changes in $f(\kappa^<)$ are relatively small. These results suggest that the small-to-large removal fraction, $\zeta^<$, does not have a strong effect in modifying the knockdown factor statistics. By contrast, in Fig.~\ref{fig:weibullstatistics}(b), when pruning the most severe imperfections with increasing $\zeta^>$, we find that the PDFs $f(\kappa^>)$ broaden considerably, shift toward higher values of $\kappa^>$, and the minimum knockdown factor $\kappa_\mathrm{min}$ (left-tail cut off of the histograms depicted by the dashed vertical dashed lines) also increases progressively with $\zeta^>$; the shells become stronger to buckling with increased $\zeta^>$. These results suggest that the knockdown factor statistics are strongly affected by the most severe defects (those with the largest amplitudes), which are progressively removed with increasing $\zeta^>$. 
These findings also point to a possible relationship, which we will investigate later in this manuscript, between the minimum knockdown factor value, $\kappa_\mathrm{min}$ and the plateau values, $\kappa_\mathrm{p}$, in the $\kappa(\bar{\delta})$ curve, for the single-defect case presented Fig.~\ref{fig:onedefect}.

The probabilistic knockdown-factor data in Fig.~\ref{fig:weibullstatistics} are for a specific, albeit representative, imperfect shell with $(\langle\overline{\delta}\rangle,\Delta\overline{\delta}) = (2.0, 1.0)$. Next, in  Sections~\ref{sec:smalltolarge} and~\ref{sec:largetosmall}, we will further investigate the effect of small-to-large ($\zeta^<$) and large-to-small ($\zeta^>$) defect removal, over a wider range of the geometric parameter space. Hereon, instead of analyzing the full Weibull statistics, we will focus on measuring the mean, $\langle \kappa \rangle$, and standard deviation, $\Delta\kappa$, of the statistical ensembles, each with $n=100$ realizations. 

\section{Removal of defects: Small-to-large}
\label{sec:smalltolarge}

In this section, we explore the effect of systematically removing increasing fractions of the less severe defects (\textit{small-to-large} amplitudes) with $\zeta^<{\in }[0.0,\, 0.9]$ on the mean knockdown factor, $\langle\kappa\rangle^<$, of several multi-defect shells designs.

\begin{figure}[h!]
    \centering
    \includegraphics[width=0.77\columnwidth]{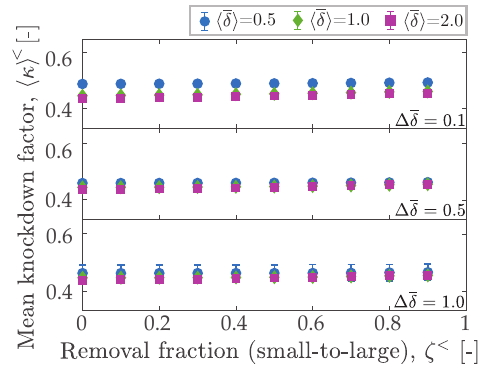}
    \caption{Mean knockdown factor, $\langle\kappa\rangle^<$, versus the small-to-large defect removal fraction, $\zeta^<$, for shells with $\overline{\delta}=\{0.5, 1.0, 2.0\}$ and $\Delta\overline{\delta}=\{0.1, 0.5, 1.0\}$. The error bars correspond to the standard deviation of $\kappa^<$ from $n=100$ statistically identical shell realizations.}
    \label{fig:smalltolarge}
\end{figure}

In Fig.~\ref{fig:smalltolarge}, we plot $\langle\kappa\rangle^<$ as a function of $\zeta^<$ for shells with $\langle\overline{\delta}\rangle=\{0.5, 1, 2\}$ and $\Delta\overline{\delta}=\{0.1, 0.5, 1\}$.
The error bars represent the standard deviation of the knockdown factor for the ensemble of $n=100$ realizations, each with the same design parameters. These error bars are relatively small; comparable or smaller than the size of the symbols in the plot. For all designs considered, we find that $\langle\kappa\rangle^<$ remains approximately constant across the full range of $\zeta^<$, indicating that removing any fraction of the smaller defects has a negligible effect (at most an increase of a few percent) on the buckling strength of these multi-defect shells.

\begin{figure*}[ht!]
    \centering
    \includegraphics[width=0.85\textwidth]{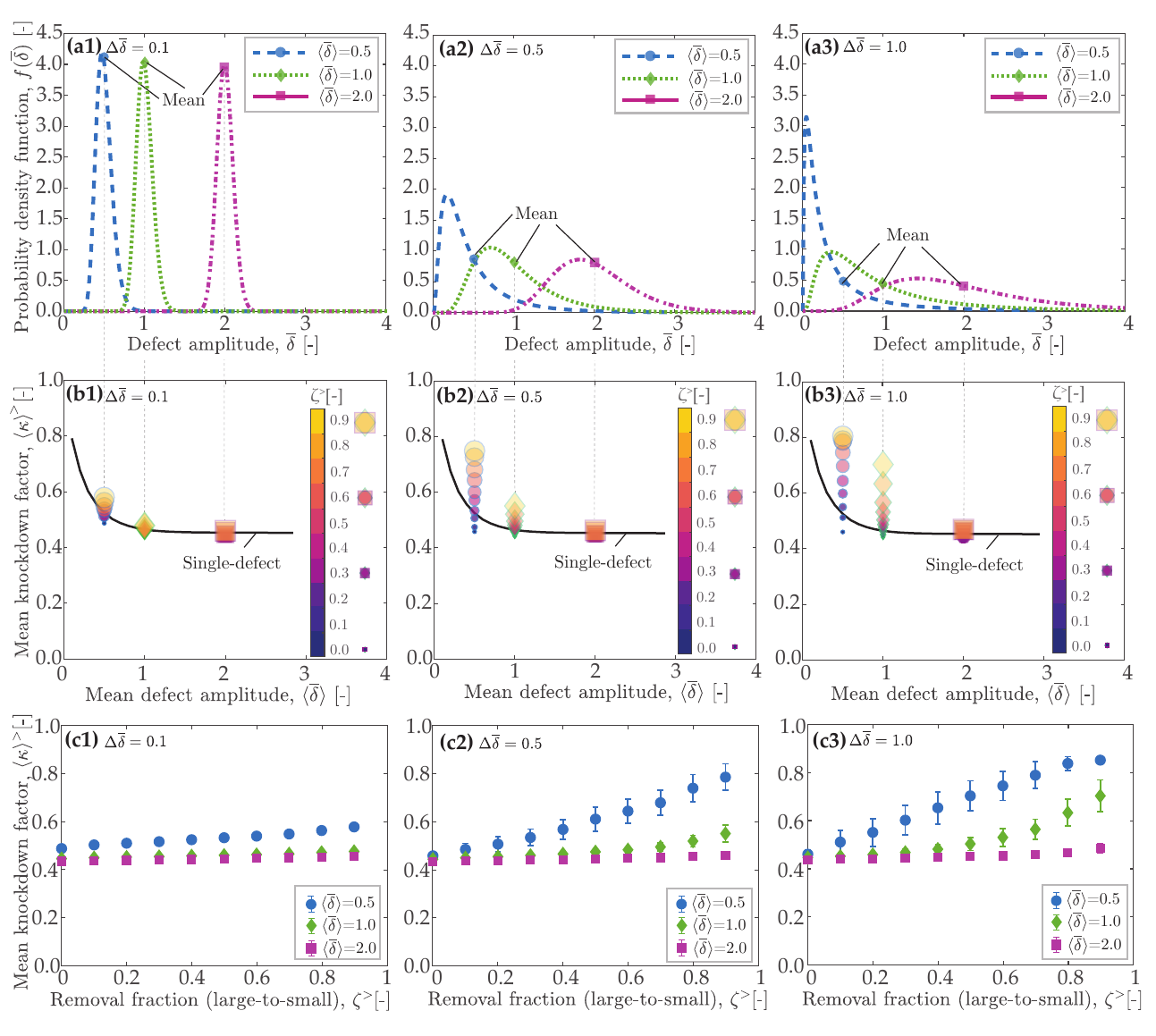}
    \caption{Knockdown factor ($\kappa^>$) statistics versus the large-to-small defect removal fraction, $\zeta^>$, for multi-defect designs characterized by $\overline{\delta}=\{0.5,\, 1.0,\, 2.0\}$ (different line types; see legend) and $\Delta\overline{\delta}=\{0.1,\, 0.5,\, 1.0\}$, in panels (abc1), (abc2), and (abc3), respectively. (a) Input probability density function (PDF), $f(\overline{\delta})$ of defect amplitude $\bar{\delta}$ used to design the multi-defect shells. The markers denote the mean defect amplitude $\langle\overline{\delta}\rangle$, which is traced by a dashed line to the respective panels, (b1), (b2), and (b2). 
    (b) Mean knockdown factor, $\langle\kappa\rangle^>$, versus mean defect amplitude, $\langle\overline{\delta}\rangle$, for various defect-removal fractions, $\zeta^>\in[0,\,0.9]$. The cases with different $\zeta^>$ are denoted by markers of increased size and color (see the adjacent colorbar). The marker symbols refer to different values of $\langle\overline{\delta}\rangle$, as indicated in the legends of panels (a). The solid black line corresponds to the knockdown factor vs. defect amplitude (with $\lambda=1$) for the single-defect case~\citep{lee2016geometric}, also reproduced in Fig.~\ref{fig:onedefect}.
    (c) Mean knockdown factor, $\langle\kappa\rangle^>$, versus $\zeta^>$;   the error bars denote the standard deviation of $\kappa^>$ for $n=100$ repetitions.
    }
    \label{fig:largetosmall}
\end{figure*}

\section{Removal of defects: large-to-small}
\label{sec:largetosmall}

We now shift to the other scenario, where we systematically remove increasing fractions of the most severe defects (\textit{large-to-small} amplitudes), $\zeta^>$.

In Fig.~\ref{fig:largetosmall}(a1,\,a2,\,a3), we present the PDFs, $f(\overline{\delta})$, with $\Delta\overline{\delta}= \{0.1,\,0.5,\,1.0\}$, respectively, used as input to generate the imperfect shell designs. The distinct line types represent different mean amplitudes (see legend), with corresponding markers denoting their positions atop the PDFs. As $\Delta\overline{\delta}$ increases, there is an increasing probability of sampling higher-amplitude defects from the right tail of $f(\overline{\delta})$, as evidenced by the increasingly broader spread of the PDFs in Figs.~\ref{fig:largetosmall}(a1,\,a2,\,a3). 

With the input designs described in the previous paragraph, in Figs.~\ref{fig:largetosmall}(b1,\,b2,\,b3), we now present the corresponding output results for the FEM-computed mean knockdown factors as a function of the mean amplitude, $\langle\overline{\delta}\rangle$, for different $\Delta\overline{\delta}$. In these plots, both the 
marker size and their color (see the adjacent color bar) represent the removal fraction $\zeta^>$. The black solid line, which is the same in the three panels, corresponds to the equivalent single-defect curve for $\lambda=1$~\citep{lee2016geometric, derveni2023probabilistic} that was already reproduced in Fig.~\ref{fig:onedefect}. In Fig.~\ref{fig:largetosmall}(b1), $\langle\kappa\rangle^>$ exhibits no increase when $\langle\overline{\delta}\rangle$ is within the plateau of the single curve, but a 19\% increase is observed for $\langle\overline{\delta}\rangle=0.5$. A more pronounced effect is found for higher values of $\Delta\overline{\delta}$ in Figs.~\ref{fig:largetosmall}(b2,\,b3), where an increase of up to 72\% and 85\% is observed for $\Delta\overline{\delta}=0.5$ (panel b2) and $\Delta\overline{\delta}=1.0$ (panel b3), respectively. 

In Fig.~\ref{fig:largetosmall}(c1,\,c2,\,c3), we replot the same datasets in panels (b1,\,b2,\,b3), this time with $\langle\kappa\rangle^>$ as a function of $\zeta^>$. The error bars represent the standard deviation of $\kappa^>$ obtained from $n=100$ statistically identical realizations. For shell designs with $\langle\overline{\delta}\rangle=2.0$, which is decidedly within the knockdown factor plateau region, $\langle\kappa\rangle^>$ remains constant with $\zeta^>$ for all values of $\Delta\overline{\delta}$ explored (each in the three panels). By contrast, for  $\langle\overline{\delta}\rangle=\{0.5,\,1.0\}$, which are before or near the onset of the plateau, respectively, there is a greater increase of $\langle\kappa\rangle^>$, and its standard deviation, with $\zeta^>$, especially for  $\Delta\overline{\delta}=0.5$ and $\Delta\overline{\delta}=1.0$. For example, in Fig.~\ref{fig:largetosmall}(c3), when 
$\langle\overline{\delta}\rangle=0.5$, the mean knockdown factor can increase by as much as 85\%, from $\langle\kappa\rangle^>=0.46$ to $\langle\kappa\rangle^>=0.85$, when $\zeta^>=0.9$ of the largest defects are removed.

The main finding emerging from Fig.~\ref{fig:largetosmall} is consistent with our prior observations~\cite{derveni2023defect,derveni2023probabilistic}, and also analogous to Carlson's experimental work~\citep{Carlson1967}: the buckling strength of multi-defect shells is governed by the most severe defects.
In our present study, where we are now addressing this problem more comprehensively, we observe that the increase in the knockdown factor is highly influenced by the design parameters of the input distribution of the defect amplitude ($\langle\overline{\delta}\rangle$, $\Delta\overline{\delta}$). The enhancement of the buckling strength, as the most severe defects are increasingly removed, is particularly remarkable for design distributions with lower mean amplitudes and higher standard deviation of the defect amplitude, as seen in the specific case with $\langle\overline{\delta}\rangle=0.5$ and $\Delta\overline{\delta}=1.0$ in Fig.~\ref{fig:largetosmall}(c3). 

\section{Comparison with single-defect shells}
\label{sec:FEMsingledefect}

In this Section, we seek to gain additional insight into the results from Section~\ref{sec:largetosmall} by comparing multi-defect shells to their single-defect shells counterparts, focusing on cases where the single defect's amplitude matches the maximum (worst) defect in the multi-defect shells characterized above.

\begin{figure}[h!]
    \centering
    \includegraphics[width=0.73\columnwidth]{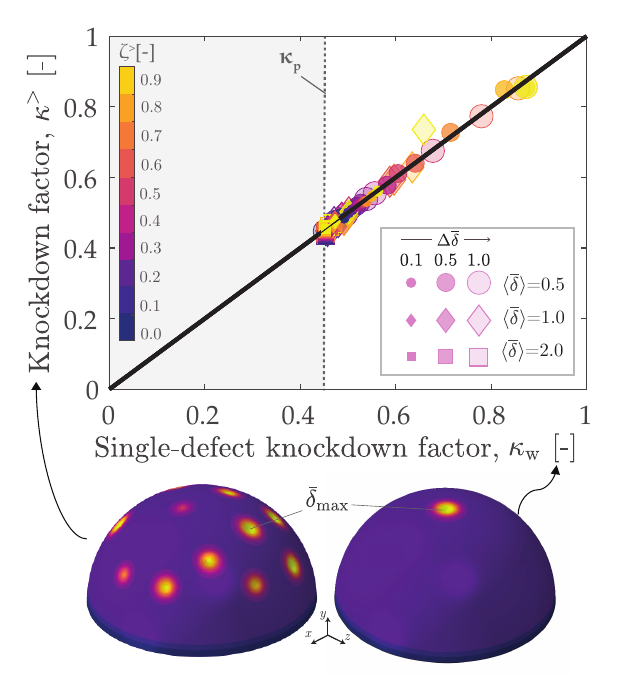}
    \caption{Knockdown factors, $\kappa^>$, for a single realization ($n=1$) of a shell containing multiple defects versus the knockdown factor, $\kappa_w$, of the equivalent shell with only one defect of the \textit{worst} amplitude from the multi-defect shell. The color map corresponds to various defect removal fractions, $\zeta^>$, and the symbols (see legend) to various geometric parameters ($\langle\overline{\delta}\rangle$, $\Delta\overline{\delta}$).
    The black solid line represents $\kappa^> = \kappa_w$, and the gray shaded region threshold corresponds to the plateau region, $\kappa_\mathrm{p}$, of the single-defect curve. Underneath the plot, we show two 3D representations of imperfect shells, one for the multi-defect case with $\zeta^>=0$ (left) and the other for the equivalent single-defect case. In both, the location of the worst defect, with $\overline{\delta}_\mathrm{max}$, is indicated.
    }
    \label{fig:singlerealization}
\end{figure}

In Fig.~\ref{fig:singlerealization}, we plot the knockdown factor $\kappa^>$ of a single realization ($n=1$) of a multi-defect shell, whose worst defect has an amplitude of $\overline{\delta}_\mathrm{max}$, as a function of the knockdown factor, $\kappa_\mathrm{w}$, of the equivalent single-defect shell, whose defect amplitude is also $\overline{\delta}_\mathrm{max}$; see the 3D visualizations underneath the plot that are representative of each case. In that plot, we superpose data for various defect removal fractions, $\zeta^>$ (see color bar), and various geometric parameters $(\langle\overline{\delta}\rangle$, $\Delta\overline{\delta})$; see different symbols in the legend). We find that all the data points collapse onto the $\kappa^> = \kappa_w$ (black) line, confirming that the worst defect with $\overline{\delta}_\mathrm{max}$ fully governs the buckling strength of these multi-defect shells. Note that no data is found to the left of the vertical dashed line (in the grey region), which corresponds to the onset of the plateau of the single-defect case (cf. Fig.~\ref{fig:onedefect}).

\begin{figure}[h!] 
    \centering
    \includegraphics[width=0.73\columnwidth]{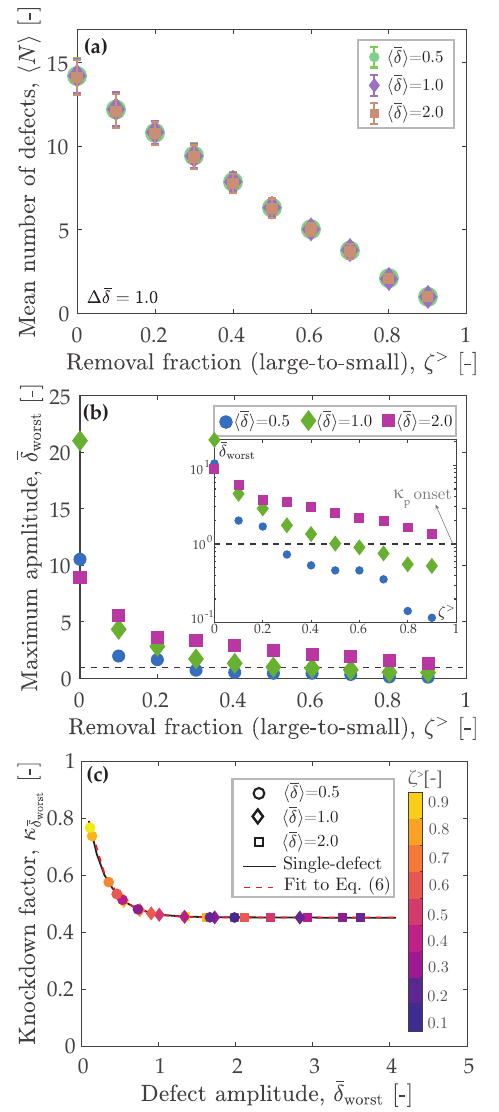}
    \caption{Statistics of the design (input) parameters of multi-defect shells ($n=100$ realizations) and comparison with single-defect cases. (a) Average number of defects, $\langle N \rangle$, vs. defect removal fraction, $\zeta^>$; the error bar refers to the standard deviation of $N$. (b) Amplitude of the worst defect, $\overline{\delta}_\mathrm{worst}$, from the entire sample of $n=100$ realizations, for three data sets with $\langle \overline{\delta} \rangle =\{0.5,\,1.0,\, 2.0\}$. (c) Fit of Eq.~(\ref{eqn:predictedknockdownfactor}) 
    (red dashed line) to the single-defect curve (black solid line).     The symbols correspond to the interpolation of the knockdown factor $\kappa_\mathrm{\overline{\delta}_\mathrm{worst}}$ by substituting 
    $\overline{\delta}_\mathrm{worst}$ from panel (b) into Eq.~(\ref{eqn:predictedknockdownfactor}). The colorbar represents the defect removal fraction. The standard deviation of the defect amplitude is fixed at $\Delta \overline{\delta}=1.0$
    }
    \label{fig:comparison}
\end{figure}

Building on the above results for an individual ($n=1$) realization, we proceed by providing evidence that the worst defects also govern the knockdown factor statistics for statistical ensembles with many realizations ($n=100$). In Fig.~\ref{fig:comparison}(a,\,b), we characterize the design (input) parameters for multi-defect shells as functions of the defect removal fraction, $\zeta^>$, specifically: the average number of defects $\langle N \rangle$ in panel (a) and the amplitude of the worst defect, $\overline{\delta}_\mathrm{worst}$ from the entire sample of $n=100$ realizations in panel (b). 
Unsurprisingly, we find that $\langle N \rangle$ decays linearly with $\zeta^>$, from $\langle N \rangle \approx14$ at $\zeta^>=0$ to $\langle N \rangle=1$ at $\zeta^>=0.9$. 
Next, in Fig.~\ref{fig:comparison}(b), we identify the amplitude of the worst defect, $\overline{\delta}_\mathrm{worst}$, from the entire statistical sample, and plot it as a function of $\zeta^>$. Three data sets with $\langle \overline{\delta} \rangle = \{0.5,\, 1.0,\, 2.0\}$ are considered (see legend). We find that for higher $\langle \overline{\delta} \rangle$ the most severe defect is still sampled from the knockdown factor plateau (above the dashed line) even for high $\zeta^>$. By contrast, for smaller $\langle \overline{\delta} \rangle$, the amplitude of the worst defect decreases much more significantly with $\zeta^>$; some of the points are in the region outside the knockdown factor plateau (below the dashed line). The semi-logarithmic plot in the inset of Fig.~\ref{fig:comparison}(b) enables a clearer visualization of the data above/below the dashed line (inside/outside the knockdown factor plateau).

We will now further explore the dominance of the worst (most severe) defect on the buckling of the multi-defect shells characterized above, seeking to establish a connection to the single-defect case in Fig.~\ref{fig:onedefect}. To do so, we first need to be able to interpolate the data in Fig.~\ref{fig:onedefect}, for any defect amplitude, by introducing the empirical function:
\begin{equation}\label{eqn:predictedknockdownfactor}
\kappa_\mathrm{\overline{\delta}_\mathrm{worst}}= \kappa_\mathrm{p} + a \exp^{-b \overline{\delta}_\mathrm{worst}},
\end{equation}
where $\kappa_p$ is the knockdown factor in the plateau of the single-defect curve (see Fig.~\ref{fig:onedefect}), $a$ and $b$ are fitting parameters, and the $\overline{\delta}$ of the single-defect case will be interpreted as, and replaced by, $\overline{\delta}_\mathrm{worst}$ in the multi-defect case. Fitting Eq.~(\ref{eqn:predictedknockdownfactor}) to the single-defect data in Fig.~\ref{fig:onedefect}
yields $a=0.490 \pm 0.003$ and $b=3.87 \pm 0.03$. In Fig.~\ref{fig:comparison}(c), we reproduce the single-defect data (black solid line), to which we juxtapose the fit to Eq.~(\ref{eqn:predictedknockdownfactor}) (red dashed line). Also, in that plot, we superpose the interpolated values for $\kappa_\mathrm{\overline{\delta}_\mathrm{worst}}$ obtained from Eq.~(\ref{eqn:predictedknockdownfactor}) for the specific values of $\overline{\delta}_\mathrm{worst}$ obtained in Fig.~\ref{fig:comparison}(b), focusing, for clarity, on all three examined data sets with $\langle \overline{\delta} \rangle = \{0.5,\,1.0,\,2.0\}$. The agreement between the individual data points, the red dashed line (fit), and the black solid line conveys the accuracy of the interpolation, which we will use next. 

\begin{figure}[h!]
    \centering
    \includegraphics[width=0.76\columnwidth]{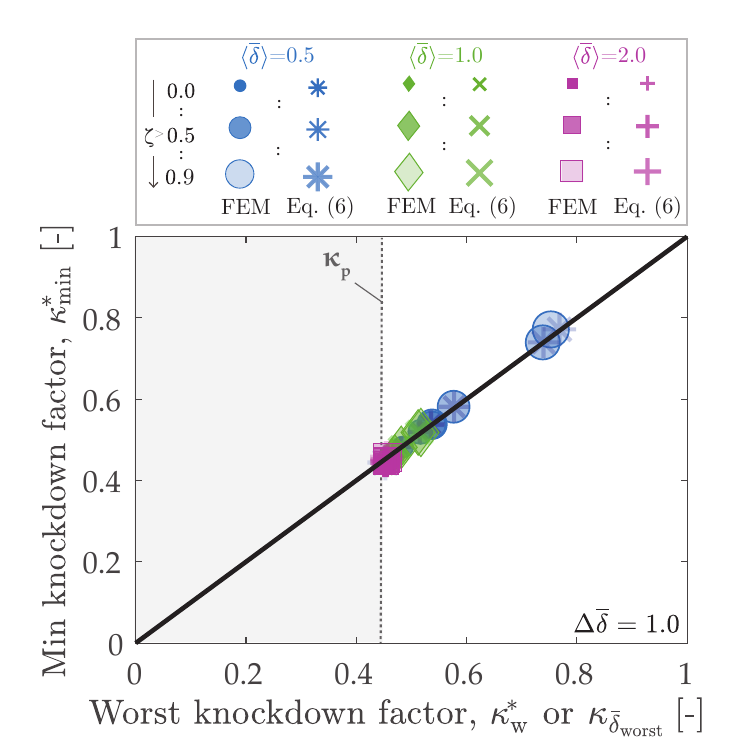}
    \caption{Minimum knockdown factor, $\kappa^*_\mathrm{min}$, vs. the knockdown factor $\kappa^*_\mathrm{w}$ or $\kappa_\mathrm{\overline{\delta}_\mathrm{worst}}$ arising, respectively, from single-defect FEM simulations with the worst amplitude from $n=100$ realizations or through interpolation using  Eq.~(\ref{eqn:predictedknockdownfactor}). The size of the markers represents the defect removal fraction, $\zeta^>$, and the black solid line represents $\kappa^*_\mathrm{min} = \kappa^*_w$ or $\kappa^*_\mathrm{min} = \kappa_\mathrm{\overline{\delta}_\mathrm{worst}}$. The gray shaded region threshold corresponds to $\kappa_\mathrm{p}$ of the plateau in the single-defect $\kappa(\delta)$ curve .}
    \label{fig:globalworstdefect}
\end{figure}

Let us now revisit the data in Fig.~\ref{fig:singlerealization}, which were computed for a \textit{single} realization ($n=1$), and attempt to relate it to the statistical case of many ($n=100$) realizations. For simplicity, we focus on the representative case of shells with $\Delta\overline{\delta}=1$. First, for this statistical ensemble (with a given input design of $\langle\overline{\delta}\rangle$), we identify the defect with maximum amplitude, the \textit{worst} defect, $\overline{\delta}_\mathrm{worst}$, amongst all of the 100 realizations, which was already plotted in Fig.~\ref{fig:comparison}(b). Then, we construct a single-defect shell with an imperfection of this amplitude, $\overline{\delta}_\mathrm{worst}$, and perform an FEM simulation to compute the corresponding knockdown factor, $\kappa^*_\mathrm{w}$.  Under the interpretation that the worst defect governs the buckling strength of multiple defects holds, we expect that $\kappa^*_\mathrm{w}$ for this single-defect shell should mimic the minimum knockdown factor $\kappa^*_\mathrm{min}$ extracted from the statistical ensemble obtained with the same input parameters; see Fig.~\ref{fig:weibullstatistics} and Section~\ref{sec:Weibull}. In Fig.~\ref{fig:globalworstdefect}, we plot the comparison between $\kappa^*_\mathrm{min}$ and $\kappa^*_\mathrm{w}$, with the black line representing $\kappa^*_\mathrm{min}=\kappa^*_\mathrm{w}$. We find that all the data with $\langle\overline{\delta}\rangle = \{0.5,\,1.0,\, 2.0\}$ (indicated, respectively, by the different solid symbols: circles, diamonds, and squares) and in the range of removal fractions $\zeta^> {\in }[0.1, 0.9]$ (indicated by the marker size) collapse onto the $\kappa^*_\mathrm{min}=\kappa^*_\mathrm{w}$ line, within deviations below 4\%. This collapse evidences that, despite the random distribution of imperfections for these shells, the most \textit{critical} (\textit{i.e.}, the lowest) knockdown factor can be accurately predicted by identifying the most severe (\textit{i.e.}, the worst) geometric defect of largest amplitude.

In Fig.~\ref{fig:globalworstdefect}, we also compare the minimum knockdown factor with the interpolated $\kappa_\mathrm{\overline{\delta}_\mathrm{worst}}$ values obtained from Eq.~(\ref{eqn:predictedknockdownfactor}), which are plotted as the open markers (stars, crosses, plus signs). Here, we know the \textit{worst} defect (\textit{i.e.}, of maximum amplitude) from the statistical ensemble, enabling us to predict the knockdown factor from the fitted single-defect curve. These data collapse onto the black line ($\kappa^*_\mathrm{min} = \kappa_\mathrm{\overline{\delta}_\mathrm{worst}}$),  and also align with the FEM-computed data (closed symbols). With these observations, we arrive at the main finding from the present study: the knockdown factor of hemispherical shells containing a distribution of defects can be predicted by identifying the worst (largest-amplitude) defect and interpolating (or computing) the buckling strength for a shell containing a single defect at its pole, of that amplitude.

\section{Conclusions} 
\label{sec:conclusions}

We employed FEM simulations, which were validated previously against experiments, to investigate the buckling strength of elastic hemispherical shells containing a random distribution of imperfections, focusing on the effect of the systematic removal of these defects. Specifically, we examined two scenarios: (a) removing defects from small-to-large amplitude, with removal fraction $\zeta^<$, and (b) removing defects from large-to-small, with removal fraction $\zeta^>$. We considered statistical ensembles (each comprising $n=100$ realizations) of multi-defect shells over a range of design (input) parameters, with non-interacting defects whose amplitudes were sampled from a lognormal distribution with a set mean and standard deviation, and with a normalized width of $\lambda=1$. We compared the buckling strength, measured by the knockdown factor obtained through FEM simulations, of these multi-defect shells with that of an equivalent shell containing a single defect of amplitude corresponding to the worst (of largest amplitude, $\overline{\delta}_\mathrm{worst}$) measured in the multi-defect case. Furthermore, we computed a knockdown factor by interpolating an empirical fitting, Eq.~(\ref{eqn:predictedknockdownfactor}), of the single-defect curve~\citep{lee2016geometric} and substituting $\overline{\delta}$ with $\overline{\delta}_\mathrm{worst}$.

We found that the most severe defect governs the buckling strength of spherical shells containing multiple imperfections, and upon its removal, the shells exhibit an increased knockdown factor by up to $0.85$. By contrast, removing the least severe defects had a negligible effect (less than $4\%$) on the knockdown factor. Our systematic results highlight the dominance of the most severe defect, as evidenced by the fact that the knockdown factor of multiple-defect shells was nearly identical to the knockdown factor of an equivalent shell containing a single defect of amplitude corresponding to the \textit{worst} defect with $\overline{\delta}_\mathrm{worst}$ of the multi-defect distribution; the single-defect knockdown factor was computed both by FEM simulations and interpolated through Eq.~(\ref{eqn:predictedknockdownfactor}). These findings demonstrate that accurate estimation of the knockdown factors of more realistic shells with random, albeit non-interacting, imperfections can be obtained from the equivalent single-defect case, whose buckling strength can even be characterized in detail, \textit{a priori}, for interpolation. 

Whereas the importance of the most severe defect (\textit{imperfection sensitivity}) of shell buckling has long been recognized in the literature, our study is the first, to the best of our knowledge, to provide a quantitative and systematic exploration of this phenomenon within a probabilistic framework for multi-defect shells, and with a wide exploration of the geometric design space. Our results also suggest a qualitative analogy to the seminal experiments by Carlson~\cite{Carlson1967} on imperfect metallic spherical shells, whose defects were progressively polished chemically. Carlson demonstrated that, through increased polishing, the knockdown factor of his shells could be increased from $\kappa\approx0.05$ to $\approx0.86$. Remarkably, in our systematic investigation using FEM simulations, we showed that the knockdown of our multi-defect shells could also be increased, from $\kappa\approx0.46$ to $\approx0.85$, by removing increasing fractions ($\zeta^>$) of imperfection with large-to-small amplitudes. We cannot establish a direct quantitative parallel between Carlson's work and ours since the experimental techniques at the time did now allow for a detailed characterization of how the chemical polishing smoothed out the defects. Still, our detailed and systematic numerical study provides additional support to his pioneering experimental observations that the severity of imperfections highly influences the buckling of spherical shells. As a caveat, it is also important to note that the physical shells in Carlson's study had a radius-to-thickness ratio from $R/t=1570$ to 2120, whereas we focused on $R/t=110$. As studied by López Jiménez et al.~\cite{jimenez_technical_2017}, the dependence of the knockdown factor on geometric parameters of the defect tends to become general and independent of $R/t$, as long as it is sufficiently large. Still, future studies should explore a wider range of $R/t$ values to further attest to the generality of the present results.

Our findings should open new avenues in the design of stronger spherical shells by calling for the control, or even removal, of the size of their most severe imperfection. Importantly, our work also suggests that the buckling strength of shells with a distribution of defects can be accurately predicted by performing a full scan (for example, through profilometry or X-ray tomography) to identify the properties of the most severe defect based solely on geometry, followed by conducting a FEM simulation with that single defect. Future research should address the extent of applicability of this approach to more complex imperfection scenarios, including through-thickness or material imperfections, interacting defects, and other shell geometries, including shallow and cylindrical shells.

\section*{Acknowledgments:}
We are grateful to John Hutchinson and Michael Gomez for fruitful discussions and suggestions.
\bibliographystyle{apssamp}
\bibliography{apssamp}

\end{document}